\documentclass[prd,aps,twocolumn,a4paper,showkeys,nofootinbib]{revtex4-1}

\usepackage{amsmath}
\usepackage{amsfonts}
\usepackage{amssymb}	
\usepackage{graphicx}
\usepackage{amsbsy}
\usepackage{color}
\usepackage{commath}
\usepackage{hyperref}

\allowdisplaybreaks

\begin{document}

\title{First ICCUB Numerical Relativity Waveform Catalog of Eccentric Black Hole Binaries}

\author{Juan \surname{Trenado}${}^{1}$}
\author{Tomas \surname{Andrade}${}^{1,2}$}
\author{Ana \surname{Climent}${}^{1}$}
\author{Maria-Antonia \surname{Ferrer}${}^{1,3}$}

\affiliation{${}^1$Departament de F{\'\i}sica Qu\`antica i Astrof\'{\i}sica, Institut de Ci\`encies del Cosmos, Universitat de Barcelona, Mart\'{\i} i Franqu\`es 1, E-08028 Barcelona, Spain, \\
${}^2$Barcelona Supercomputing Center (BSC), CASE Department, C/ Jordi Girona 29, Nexus II Building, Barcelona, Spain, \\
${}^3$Departament de F\'isica, Universitat de les Illes Balears, IAC3--IEEC, E-07122 Palma, Spain}

\begin{abstract}
We release the first Numerical Relativity catalog of Institut de Ciencies del Cosmos at University of Barcelona (ICCUB) consisting of 128 simulations for black hole binaries. All simulations in this first release correspond to highly eccentric binaries with eccentricity $e = (0.62,0.79)$ which develop zoom-whirls up to three close passages before merger. We consider aligned, equal spin configurations in the range $\chi = (-0.5, 0.5)$ and equal mass ratios.
For each simulation, we provide the modes $(\ell, m)$ of Weyl scalar $\psi_4^{(\ell,m)}$ extrapolated to $r = \infty$, with $\ell \leq 4$. In addition, we provide the corresponding strain modes  obtained by computing a double time integral of the Weyl scalar modes. Moreover, we provide metadata and the parameter files required to reproduce our results using the open-source code Einstein Toolkit. 
A Python code that facilitates the access to the data is available on Git-Hub. 

\end{abstract}

\date{\today}

\maketitle

\section{Introduction}

The groundbreaking detection of gravitational waves by the LIGO-Virgo-KAGRA collaboration in 2015 marked the dawn of a new era in astrophysics \cite{LIGOScientific:2016aoc}. This achievement is the culmination of a multidisciplinary effort, combining the exceptional precision of advanced detectors with sophisticated data analysis techniques. A crucial element underpinning these efforts is the accurate theoretical modeling of gravitational waves \cite{Jaranowski:2005hz}. 

The most precise approach for modeling black hole binaries is through full numerical simulations of General Relativity, a field known as Numerical Relativity (NR). Since its inception, NR has evolved into a well-established discipline. The first successful simulations of binary black hole (BBH) systems in 2005 \cite{Pretorius:2005gq, Campanelli:2005dd, Baker:2005vv} opened the door to significant advancements, leading to the development of a variety of robust codes \cite{spec, ETK, Daszuta:2021ecf, Andrade:2021rbd} that now enable researchers to perform these simulations routinely. 

To date, the majority of NR simulations have focused on the quasi-circular regime, which is characteristic of BBHs formed in isolation. However, recent studies have highlighted the importance of eccentricity as a distinctive signature of BBH formation in dense astrophysical environments \cite{East:2012xq, Rasskazov:2019gjw, Tagawa:2019osr}. Modeling such systems requires NR simulations that extend beyond the quasi-circular paradigm. 

In this first release, we concentrate on the relatively unexplored domain of highly eccentric dynamical captures. These events are physically motivated by the emission of gravitational radiation during one or more close encounters between black holes, ultimately leading to their merger. 

The primary NR databases available today include the SXS~\cite{Mroue:2013xna,Chu:2015kft,Boyle:2019kee}, RIT~\cite{Healy:2017psd, Healy:2019jyf, Healy:2020vre, Healy:2022wdn}, CoRe~\cite{Dietrich:2018phi, Gonzalez:2022mgo}, SACRA-MPI \cite{Kiuchi:2017pte,Kiuchi:2019kzt}, Cardiff~\cite{Hamilton:2023qkv}, and MAYA~\cite{Ferguson:2023vta} catalogs. While most of these databases focus on simulations of quasi-circular, low-eccentricity BBHs, dynamical capture simulations have only been added to the latest RIT  \cite{Healy:2022wdn} and SXS \cite{scheel2025sxscollaborationscatalogbinary} releases.

Our simulations explore similar phenomenology to those found in the RIT catalog and cover a comparable parameter space. Specifically, as in \cite{Healy:2022wdn}, the initial energies considered in our simulations are less than the total mass, ensuring that the systems we model are bound. This approach allows us to accurately represent the later stages of dynamical captures, as argued in \cite{Andrade:2023trh}\footnote{NR simulations of dynamical captures with truly unbound initial date have been recently considered in \cite{Albanesi:2024xus}}.
However, an important methodological distinction in our work is the use of initial data with a larger separation between the black holes, which helps mitigate certain numerical artifacts. Previous studies of dynamical capture NR simulations can be found in \cite{Gold:2012tk, East:2012xq, Nelson:2019czq, Bae:2020hla, Gamba:2021gap, Albanesi:2024xus, Nagar:2020xsk, Andrade:2023trh, Carullo:2023kvj}. 

This paper accompanies the first release of the catalog from the ICCUB Numerical Relativity group. The data includes 128 simulations of equal-mass, spinning BBH systems undergoing highly eccentric dynamical captures. These simulations exhibit "zoom-whirl" behavior, with up to three close encounters before merger. The black hole spins are aligned with the orbital angular momentum and equal to each other, with spin parameters in the range $\chi =(-0.5, 0.5)$. 

We extract the Weyl scalar $\psi_4^{(\ell,m)}$ modes for $\ell \leq 4$ at several finite radii and extrapolate to $r = \infty$ using a first-order extrapolation method \cite{Nakano:2015pta}. Based on the definition of eccentricity introduced in NR simulations in \cite{Healy:2017zqj, Healy:2022wdn}, the eccentricity of our initial configurations spans the range $e = (0.62,0.79)$.

The NR data is available at \cite{ECatalogue:2025}. The Python code to explore the data and experiment with other choices of the post-processing procedure is at \cite{ECatalogue_github:2025}.

This article is organized as follows. In Section \ref{sec:num_methods} we
briefly summarize the methods for producing the numerical simulations and for post-processing the output data. In Sec. \ref{sec:catalog} we describe the content of the data in the catalog. Sec. \ref{sec:concl} concludes with plans to expand simulations and enhance tools for improved waveform modeling in future work.

\section{Numerical Methods}
\label{sec:num_methods}
\subsection{Time evolution}

We have carried out a set of 128 NR simulations using the open source code \texttt{EinsteinToolkit} \cite{ETK}. Our setup is as follows.
We use the thorn \texttt{TwoPunctures}~\cite{Brandt:1997tf, Ansorg:2004ds} to solve for the initial data. 
We consider two black holes of individual (puncture) ADM masses $M_{1,2}$, with the total mass of the simulation given by $M = M_1 + M_2$. We introduce rotation in the initial black holes aligned with the orbital angular momentum as $\vec J_{1, 2} = \chi M_{1, 2}^2 (0, 0, 1)$, with $\chi$ the rotation parameter of each black hole. The initial coordinate separation for all our simulations is $D = 40M$. 

For time-evolution, we employ the thorn \texttt{MLBSSN}, which implements the BSSN formulation of General Relativity \cite{Baumgarte:1998te, Shibata:1995we}. 
We use 8th order finite differences for spatial derivatives and a method-of-lines time integration with a 4th order Runge-Kutta scheme. 
We employ Kreiss-Oliger dissipation of order 9 scaled with a factor of $0.15$, and a CFL factor of $0.075$.

We use the thorn \texttt{Coordinates} which implements the multi-patch grid structure \texttt{Thornburg04}, corresponding to a cubic grid of size $1.37M$ in the region where the black holes move and a spherical grid of size $1000 M - 1500 M$ in the outer region where wave extraction is carried out. We have adjusted the size of the outer domain depending on the duration of each simulation to avoid boundary effects. 
The inner cubic region has 7 levels, with puncture resolution of $dx = 0.0214 M$ and an outermost level of $dx = 1.37 M$. The spherical region has constant radial resolution of $h = 1.37M$ and angular resolution of $n = 28$ points.  
Since the spins are always aligned with the orbital angular momentum, we employ z-reflection symmetry to reduce the size of the computational domain by one half. This can be easily implemented via the thorn \texttt{Coordinates}. 
This setting can be found in \cite{wardell_2016_155394} which we have adapted to our problem. 

We use the thorn \texttt{AHFinder} to track the locations of the moving punctures, as well as that of the final black hole formed at the center of the computational domain. Given this data, the thorn \texttt{QuasiLocalMeasures} allows us to extract the quasi-local quantities associated to mass and spin. Note that these are only well-defined in nearly steady state and when the black holes are separated from one another. 

We extract the wave content of our simulations using the thorns \texttt{WeylScal4} and \texttt{Multipole} which output the Weyl scalar $\psi_4$ expanded in spherical harmonics up to $\ell = 4$, at fixed radii $R = 100, 115, 136, 167 M$. 
%
%We have checked that $R \psi_4$ changes by less than $XXX \%$  as we vary $R$ in $R1, R2, ... M$, which are all located in a refinement level with resolution $dx = 4/2^2M = 1M$. 

\subsection{Post-processing}
\label{sec:post}

Our simulation output requires some post-processing in order to be usable for waveform modelling. First, since we extract $\psi_4$ at finite radius, we need to extrapolate to $r = \infty$ to get a fully gauge-invariant result. 
To this end, we adopt the method of first order extrapolation put forward in \cite{Nakano:2015pta}. This amounts to evaluating
\begin{multline}
\label{eq:RIT_formula}
    \left. r\,\psi_4^{\ell m}\right|_{r=\infty} = 
\alpha(r) \left( r\psi_{4\ell m}^{\rm NR}(t,r) \right. \\
\left. - \beta(r,\ell) \int dt [r\psi_{4\ell m}^{\rm NR}(t,r)] \right) \, ,
\end{multline}
where
\begin{equation}
\alpha(r) = 1-{\frac {2M}{{r}}} \quad \text{and} \quad \beta(r,\ell) = \frac{(\ell -1)(\ell +2)}{2\,r} \, .
\end{equation}
Notice the presence of the integral in \eqref{eq:RIT_formula}. Moreover, since the strain modes are given by 
\begin{equation}
    \psi_4 = \partial_t^2 h
\end{equation}
\noindent we need to perform further a double time integral to extract the strain. 
A popular method to evaluate these time integrals is the fixed-frequency integration (FFI) method \cite{Reisswig:2010di}, which is defined by 
\begin{equation}
\label{ffi}
    2\pi \mathcal{F} \left[ \int_{-\infty}^{t} h(t')dt'\right] = \left\{ -i\Tilde{h}(\omega)/\omega_0 \, , \quad \omega\leq \omega_0 \, , \atop -i\Tilde{h}(\omega)/\omega \, , \quad \omega > \omega_0 \, , \right. 
\end{equation}
where $\mathcal{F}$ is the Fourier transform, $\Tilde{h}(\omega)$ is the strain in the frequency domain $\omega$ and $\omega_0$ is set to avoid divergences of the low-frequency modes. In general, fixing $\omega_0$ is challenging: a small value can amplify unphysical low-frequency modes, while a large value may suppress important physical frequencies of the waveform. 

For quasi-circular binaries, there is a natural choice for $\omega_0$ since the frequency increases monotonically. For dynamical captures this is no longer the case, so we proceed by selecting the value of $\omega_0$ as the one that optimizes the energy and angular momentum balance for each simulation\footnote{In \cite{Andrade:2023trh} the method of direct time integration was employed to treat dynamical capture simulations similar to the ones in this catalog. However, this method gives large drifts if the number of encounters is larger than two.}. 
%
%\twocolumn

More concretely, we consider the difference between the initial and final energy $\Delta E = E_{{\rm ADM}} - E_f$ and angular momentum $\Delta J = J_{{\rm ADM}} - J_f$, and compare these with the corresponding integrated fluxes\footnote{We sum over all modes available $\ell \leq 4$, but discard the $m = 0$ modes which are noisy. Note however that these do not enter in the angular momentum radiation.}, 
\begin{align}
    E_{{\rm rad}} &= \frac{1}{16 \pi} \sum_{(\ell, m)} \int_{t_i}^{t_f} \lvert \dot h_{\ell m} \rvert^2 \\
%    E_{{\rm rad}} &= \frac{1}{16 \pi} \sum_{(\ell, m)} \int_{t_i}^{t_f} \dot h_{\ell m} \dot h_{\ell m} \\
%
    J_{{\rm rad}} &= \frac{1}{16 \pi} \sum_{(\ell, m)} m \int_{t_i}^{t_f}  \Im [h_{\ell m} \dot h^*_{\ell m}]
\end{align}
For each simulation, we select the value of $\omega_0$ which minimizes the quantities $\delta_E = E_{{\rm ADM}} - E_f - E_{{\rm rad}}$ and $\delta_J = J_{{\rm ADM}} - J_f - J_{{\rm rad}}$. Following these procedure, we find values of $\omega_0$ which yield a $2.0\%$ and $3.5\%$ maximum difference for $\delta_E$ and $\delta _J$, respectively.

We extract the mass and spin of the final black hole around $t \approx 400 M$ after the final black hole has formed. We have checked that extracting the average of these quantities over the window $ t \approx (100 M, 500M)$
only incurs in differences of order $0.04 \%$ and $0.2 \%$ for the final mass and spin, respectively.

\section{Catalog}
\label{sec:catalog}
Our catalog of NR simulations is available at \cite{ECatalogue:2025}. The user interface provides a simple tool to filter the simulations of interest via the most common initial and post-merger parameters. 
For each simulation, we provide the $\psi_4$ multipoles for $\ell \leq 4$ extrapolated to infinity, the corresponding strain modes, a metadata file including the most relevant initial data parameters and post-merger quantities, and the Einstein Toolkit parameter file. For a full discussion on the convergence tests, see Appendix A.

Our coverage of parameter space is designed to explore the transition between different number of encounters, up to three. To do this, we fix the energy and orbital angular momentum, and vary the spin of the black holes from positive to negative values. 
We observe that increasing the total angular momentum for fixed energy leads to more encounters, in agreement with previous studies \cite{Gamba:2021gap, Placidi:2021rkh, Albanesi:2024xus}.
We expect that increasing the total angular momentum further should lead to open, scattering orbits.

Although it is well-known that eccentricity is not a gauge-invariant quantity, it is useful to estimate it to characterize NR simulations, see e.g. \cite{Knee:2022hth} for a recent discussion on this topic. We use the notion of eccentricity defined in \cite{Healy:2017zqj, Healy:2022wdn}, which uses the third post-Newtonian order methods in order to determine quasi-circular orbits. In particular, we will be interesting in computing the tangential momentum $P_{t,\text{qc}}$. For eccentric orbits, we modify the initial tangential momentum of the orbit by the factor $(1-\varepsilon)$, with $0<\varepsilon<1$, such that $P_t = (1-\varepsilon)P_{t,\text{qc}}$. The initial eccentricity $e$ is related to this factor by $e=\varepsilon(2-\varepsilon)$. 
%\AC{Shall I put more details?} \AC{In our simulations...}

\begin{figure}[] 
\centering
\includegraphics[width=0.45\textwidth]{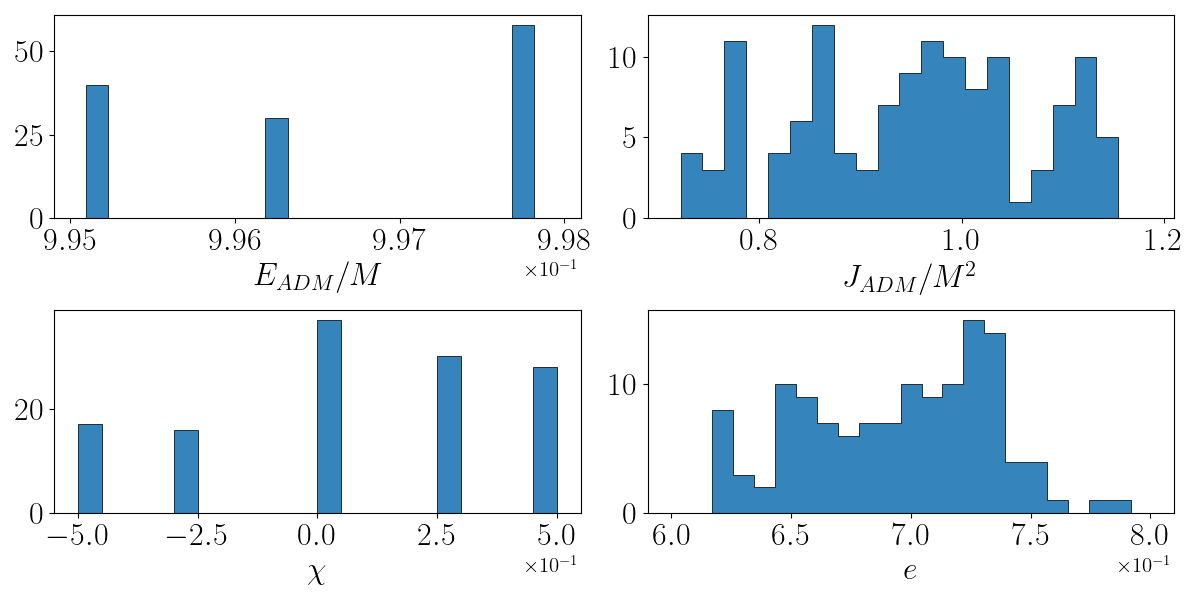}
\caption{Distributions of initial parameters $\rm (J_{ADM}, E_{ADM}, \chi)$ and resulting eccentricity, $e$.} 
\label{fig:hist}
\end{figure}

\begin{figure}[] 
\centering
\includegraphics[width=0.48\textwidth]{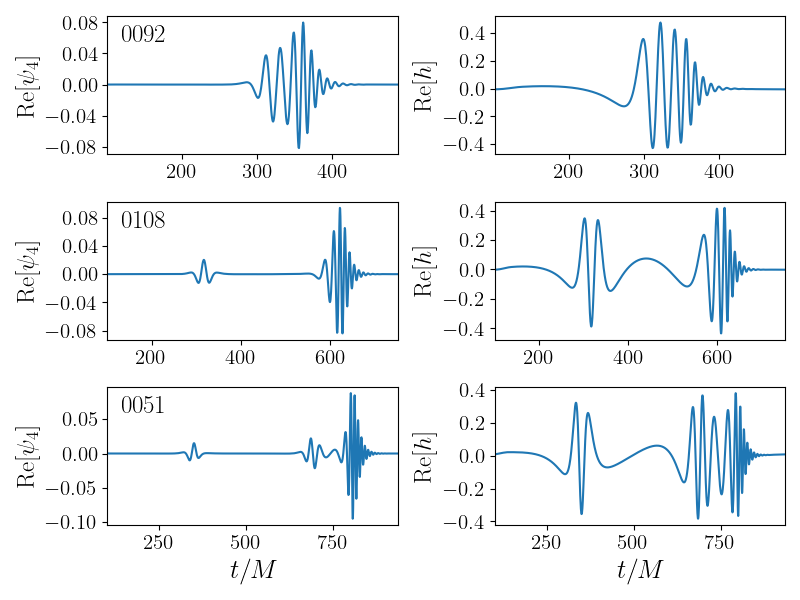}
\caption{Sample waveforms displaying one, two and three encounters corresponding to our configuraitons with IDs 0092, 0108 and 0051. The encounters are characterized by bursts of radiation which take place before the merger/ringdown. 
On the left column we show the leading mode $(2,2)$ of the Weyl scalar $\psi^4_{(2,2)}$, while on the right we show the corresponding strain. This is obtained by integrating $\psi^4_{(2,2)}$ twice using the FFI method as described in the main text.} 
\label{fig:sample_waveforms}
\end{figure}

\begin{figure*}[!htp] 
\centering
\includegraphics[width=0.8\textwidth]{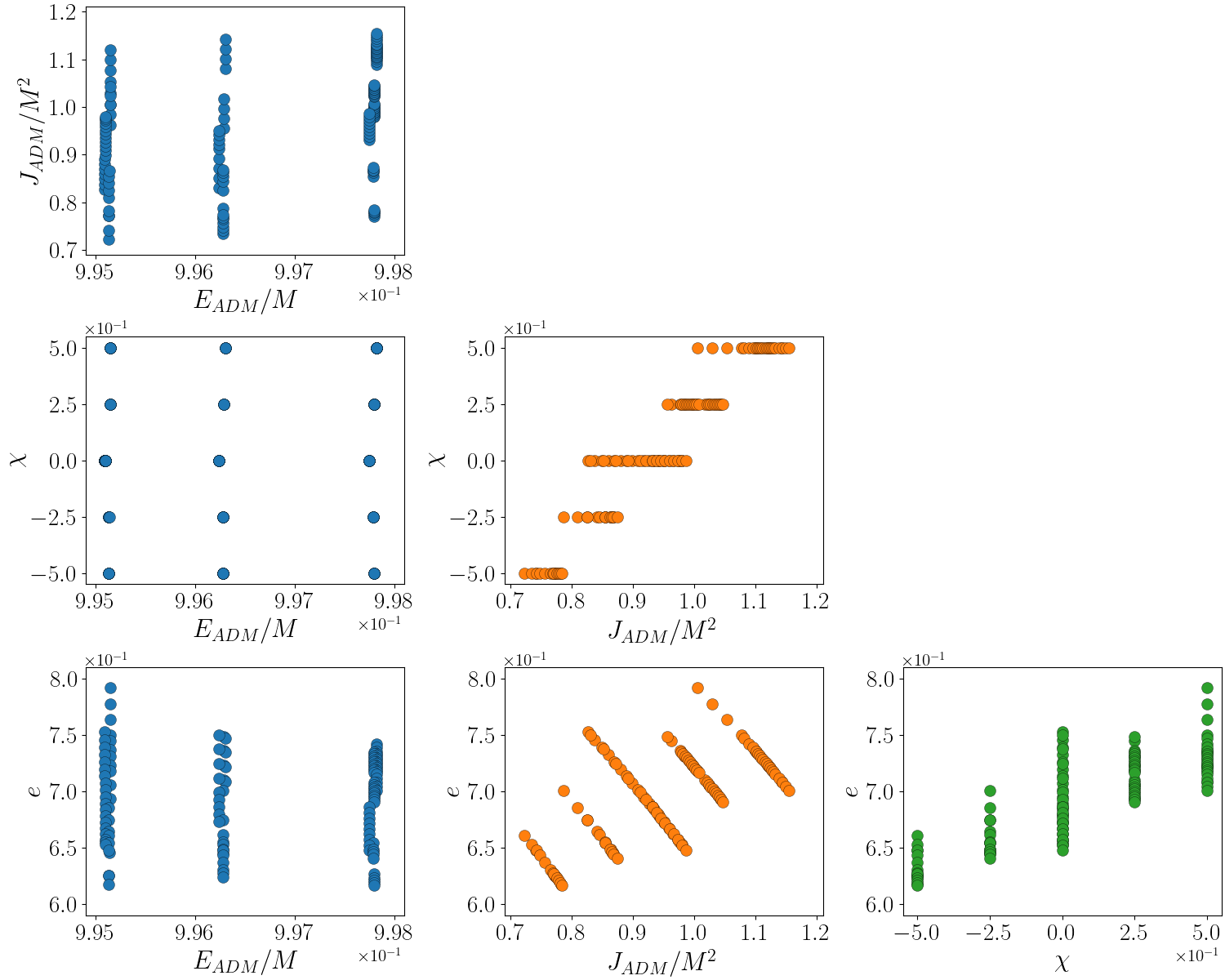}
\caption{Initial parameters for the 128 simulations $(J_{ADM}, E_{ADM}, \chi)$. We consider BBH configurations with the same aligned spin. $J_{ADM}$ and $E_{ADM}$ values have been obtained using the \texttt{TwoPunctures} thorn and eccentricity, $e$, computed following \cite{Healy:2017zqj}, is distributed over the range of $e = (0.62,0.79)$ .} 
\label{fig:param_space}
\end{figure*}

%, which could also occur when $E \lesssim M$.
%
We show different slices of our parameter space coverage in Figs \ref{fig:hist} and \ref{fig:param_space}, , and a selected sample of the waveforms in Fig. \ref{fig:sample_waveforms}. This shows the distinctive features of dynamical captures or zoom-whirls where each encounter corresponds to a burst of radiation until the final merger takes place.

\begin{figure}[] 
\centering
\includegraphics[width=0.35\textwidth]{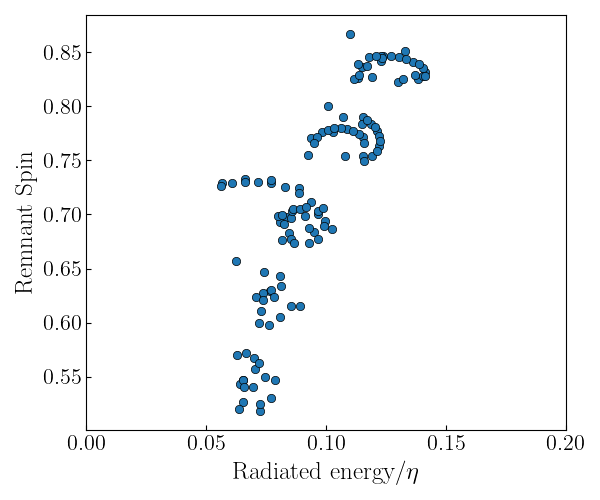}
\caption{Correlation between the remnant spin and the radiated energy, where the radiated energy is divided by the symmetric mass, $\eta = q / (1+q)^2$, to obtain the leading mass dependence.} 
\label{fig:postmerger}
\end{figure}

Fig. \ref{fig:postmerger} shows information about the final states, displaying the spin of the final black holes as a function of the total radiated energy.  

\section{Conclusions}
\label{sec:concl}

We have released 128 NR simulations for highly eccentric black hole binaries, which are a critical tool for waveform modelling of these events. 
As an unambiguous post-processing procedure of the corresponding waveforms is not yet available, we provide open source post-processing tools in which we implement our preferred choices, but give the user the flexibility to explore other alternatives. 

Interesting avenues for future improvement are  considering unequal mass binaries, increasing the number of higher modes and including multiple resolutions for our simulations. 
This will in turn require the implementation of additional post-processing tools which become more important in these regimes such as higher order extrapolation, and accounting for center of mass motion. 
We hope to incrementally address these issues in future releases.

\section{Acknowledgements}

The work of JT and TA is supported in part by the Ministry of Science and Innovation (EUR2020-112157, PID2021-125485NB-C22, CEX2019- 000918-M funded by MCIN/AEI/10.13039/501100011033), and by AGAUR (SGR-2021-01069). The work of AC is supported by MICIN through grant PRE2021-098495 and by the State Research Agency of MICIN through the ‘Unit of Excellence Maria de Maeztu 2020-2023’ award to the Institute of Cosmos Sciences (CEX2019-000918-M). The work of MAF was supported by the yearly plan of the Tourist Stay Tax for 2023 (ITS2023-086 - Programa de Fomento de la Investigación); the Universitat de les Illes Balears (UIB); the Spanish Agencia Estatal de Investigación grants CNS2022-135440, PID2022-138626NB-I00, RED2024-153978-E, RED2024-153735-E, funded by MICIU/AEI/10.13039/501100011033, the European Union NextGenerationEU/PRTR, and the ERDF/EU; and the Comunitat Autònoma de les Illes Balears through the Conselleria d'Educació i Universitats with funds from the European Union - NextGenerationEU/PRTR-C17.I1 (SINCO2022/6719) and from the European Union - European Regional Development Fund (ERDF) (SINCO2022/18146).

\bibliography{refs_local.bib, refs.bib}

\appendix
\section{Consistency checks}

\begin{figure}[] 
\centering
%\begin{minipage}{0.48\textwidth}
    \centering
    \includegraphics[width=0.4\textwidth]{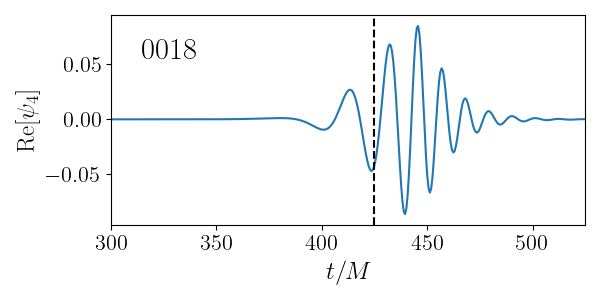}
    \includegraphics[width=0.4\textwidth]{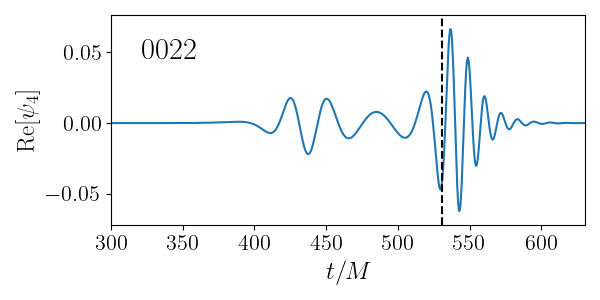}
    \includegraphics[width=0.4\textwidth]{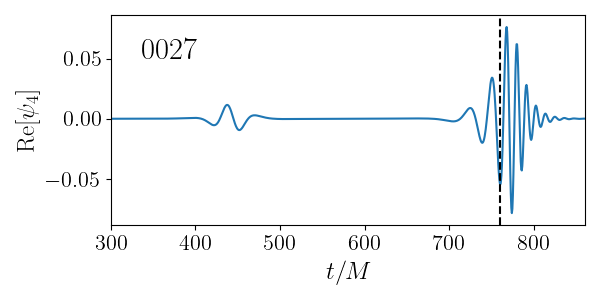}
    \caption{Real part of the $\psi_4$ scalar for the configurations used in our convergence tests, numbered 0018, 0022, 0027. The vertical dashed line shows the point of maximum amplitude of the strain of the leading mode, which can be interpreted as the merger time.} 
    \label{fig:conv_test_sample}
%\end{minipage}
\end{figure}

We have carried out a limited number of simulations at various resolutions to explore the convergence of our methods, see Fig \ref{fig:conv_test_sample} for the
depiction of the real part of the leading modes.  
These correspond to configurations 0018, 0022, 0027 of our main dataset. The parameters for the simulations included in the convergence tests can be found in Table \ref{tab:self_conv}. 

\begin{table}[h]
\caption{\label{tab:self_conv} [Grid settings for the convergence tests of the configurations 0018, 0022, 0027. We show the grid spacing at the coarsest level of the inner cubic box $\delta x$, the radial grid spacing at the extraction zone $h$ (which is held constant in the whole outer spherical grid), and the resolution at puncture $\delta x_p$, for the three resolutions Low (L), Medium (M) and High (H). All quantities are expressed in units of the total mass $M$.]}
\begin{center}
\begin{ruledtabular}
\begin{tabular}{c | c c c c c c} 
res & $\delta x$ & $h$ & $\delta x_p [10^{-2}]$  \\
\hline
L & 1.37 & 1.37  & 2.14   \\
M & 0.87 & 0.87 & 1.36  \\
H & 0.69 & 0.69 & 1.08  \\

\end{tabular}
\end{ruledtabular}
\end{center}
\end{table}

%\hfill

As customary, we separate the waveforms into amplitude and phases according to 
\begin{align}
    \psi^4_{\ell m}(t) &= a_{\ell m}(t) e^{- i \phi_{\ell m}(t)} \\
    h_{\ell m}(t) &= A_{\ell m}(t) e^{- i \Phi_{\ell m}(t)}
\end{align}
\noindent After performing this decomposition, we evaluate convergence by comparing the time-dependent amplitude and phases of the leading modes for consecutive resolutions.

Since our grid is multi-patched, we follow  \cite{Pollney:2009yz} to carry out the convergence tests. In particular, we define the scaling factor for order $r$ as 
\begin{equation}
    SF(r) = \frac{h_H^r - h_M^r}{h_M^r - h_L^r}
\end{equation}
\noindent where $h_i$ is the radial resolution of the spherical grid for a Low, Medium and High resolutions. 
In order to check convergence for order $r$, we should have
\begin{equation}
    \frac{A_H - A_M}{A_L - A_M} \approx SF(r)
\end{equation}
\noindent and similarly for the phases.

\begin{figure}[] 
    \begin{minipage}{0.48\textwidth}
        \centering
        \includegraphics[width=0.9\textwidth]{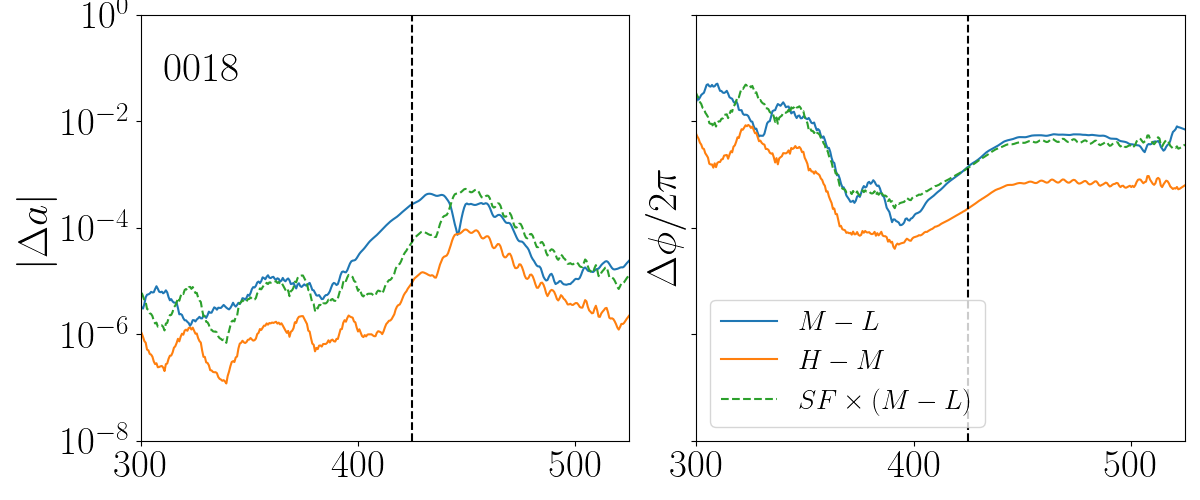}
        \includegraphics[width=0.9\textwidth]{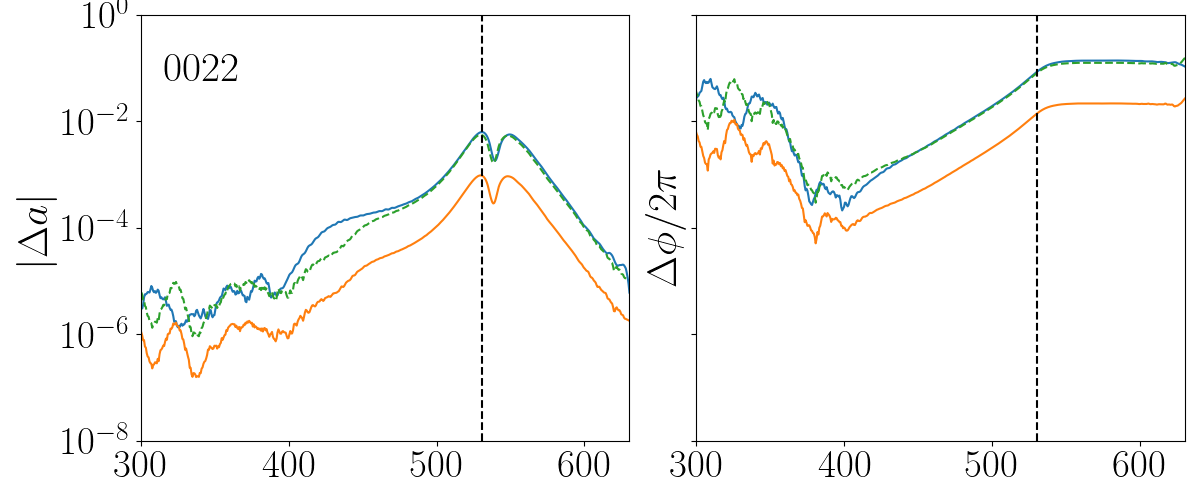}
        \includegraphics[width=0.9\textwidth]{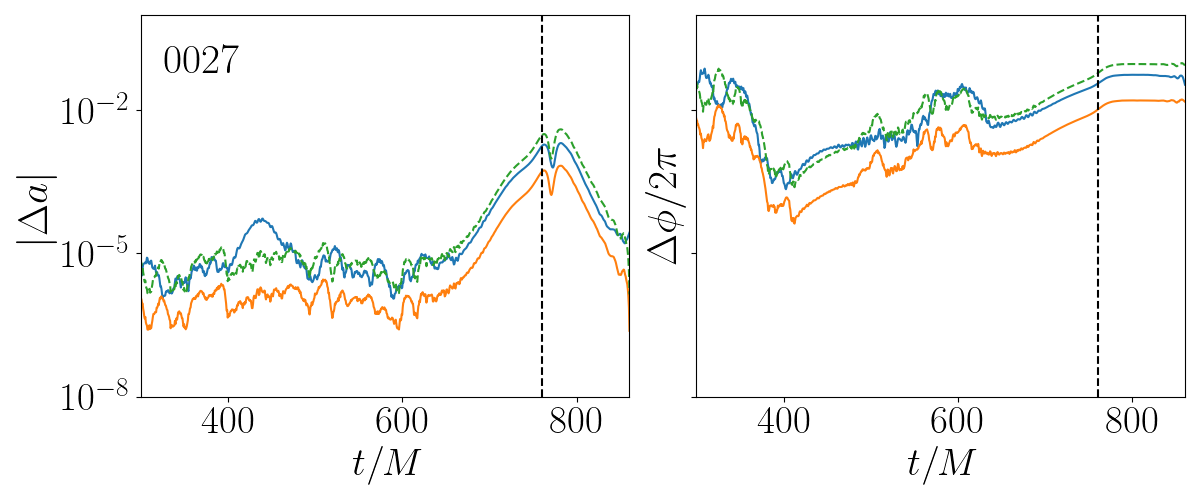}
        \caption{Convergence tests for the leading modes of the scalar $\psi_4$ for configurations 0018, 0022, 0027. We show the differences for amplitude (left) and phase (right) for consecutive resolutions in solid lines. The dashed line marks the product of the scale factor multiplied by the medium-high differences. As explained in this appendix, coincidence of this line with the medium-low resolution lines shows good convergence of order $r = 3$.} 
        \label{fig:conv_test_psi4}
    \end{minipage}
    \vspace{10cm}
    \begin{minipage}[t]{0.48\textwidth}
        \centering
        \includegraphics[width=0.9\textwidth]{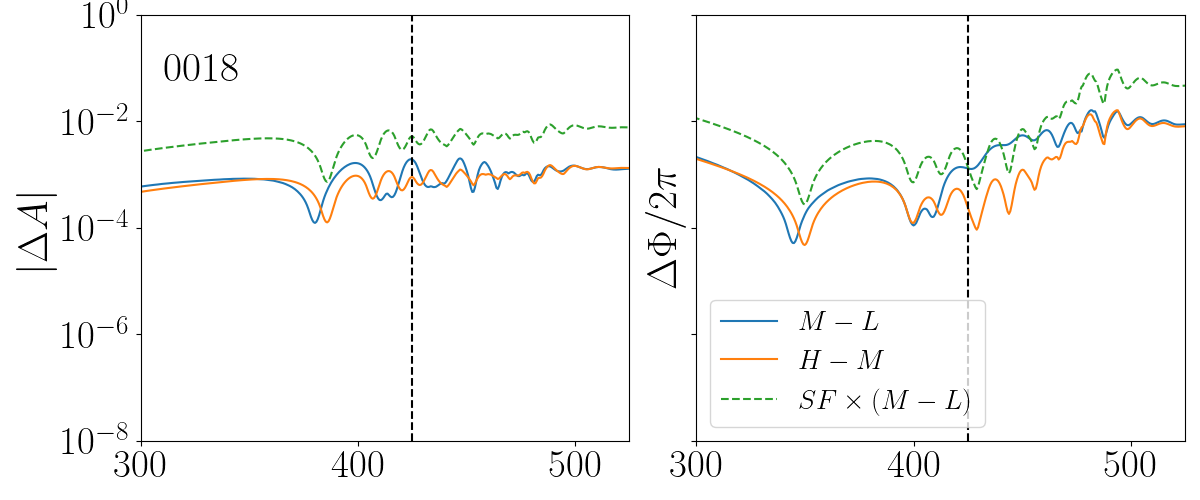}
        \includegraphics[width=0.9\textwidth]{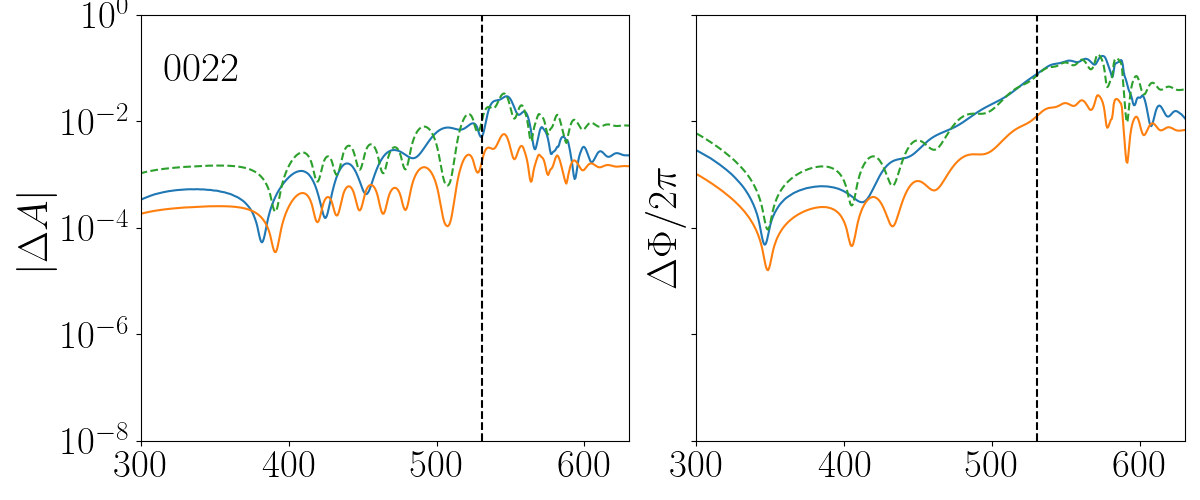}
        \includegraphics[width=0.9\textwidth]{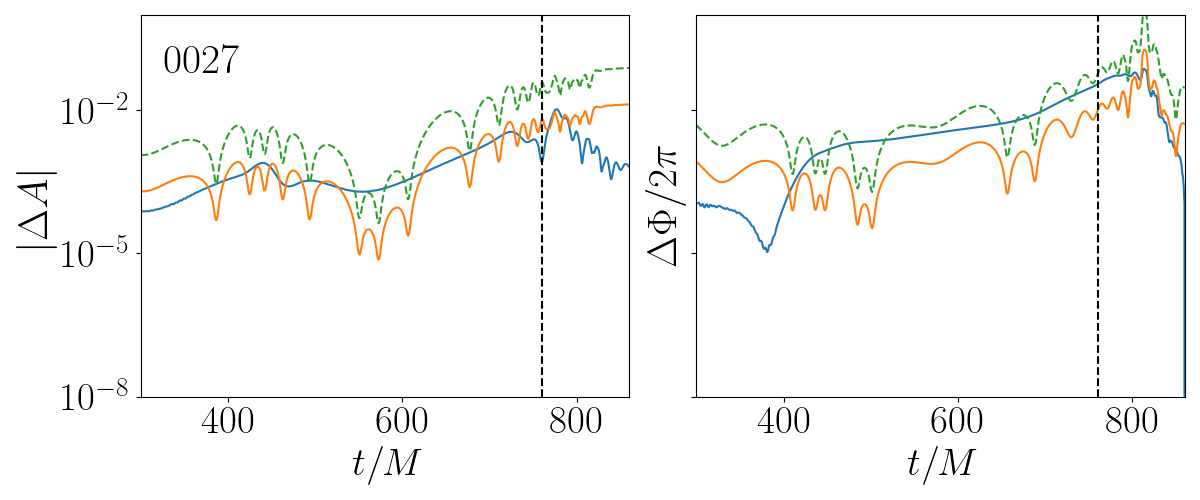}
        \caption{Convergence tests for the leading modes of the strain for configurations 0018, 0022, 0027. We show the differences for amplitude (left) and phase (right) for consecutive resolutions in solid lines. The dashed line marks the product of the scale factor multiplied by the medium-high differences. As explained in this appendix, coincidence of this line with the medium-low resolution lines shows good convergence of order $r = 3$.} 
        \label{fig:conv_test_h}
    \end{minipage}
\end{figure}

We find that the raw data $\psi^4_{(2,2)}$ extracted at radius $R = 100 M$ is approximately compatible with convergence of order 3, see Fig. \ref{fig:conv_test_psi4}. 
While differences between consecutive resolutions remain of order $10^{-4}$ after reconstructing the extrapolated strain, we observe that convergence is degraded.
In particular, for short waveforms there is a saturation at medium resolution, e.g. so no accuracy gain is obtained increasing resolution further, see Fig \ref{fig:conv_test_h}.

\end{document}